\documentclass{PoS}

\title{Heavy quarkonium production in the Regge limit of QCD: from Tevatron to LHC}

\ShortTitle{Heavy quarkonium production in the Regge limit of QCD}

\author{\speaker{Maxim Nefedov}
        \thanks{Supported by the RFBR, 12-02-31701-mol-a}\\
        Samara State University\\
        E-mail: \email{nefedovma@gmail.com}}

\author{Vladimir Saleev
\thanks{Supported by the Russian Ministry of Science and Education, Contract 14.B37.21.1182}\\
       Samara State University\\
        E-mail: \email{saleev@samsu.ru}}

\author{Alexandra Shipilova
\thanks{Supported by the RFBR, 12-02-31701-mol-a}\\
       Samara State University\\
        E-mail: \email{alexshipilova@samsu.ru}}

\abstract{Heavy quarkonium production in the framework of the
non-relativistic quantum chromodynamics and leading order of the
parton Reggeization approach at the Tevatron and LHC is discussed.
 In this note, we compare our predictions for the bottomonium production at the LHC due
 to the color-singlet approximation of the non-relativistic quantum chromodynamics with CMS
  and LHCb data. It is found, that in the production of $\Upsilon(1S)$ states,
  the color-singlet mechanism is dominating,
  whereas to describe the data for the inclusive $\Upsilon(2S)$ and $\Upsilon(3S)$ production,
  the color-octet contributions should be
  taken into account.}

\FullConference{Xth Quark Confinement and the Hadron Spectrum\\
                 8-12 October 2012\\
                 TUM Campus Garching, Munich, Germany}

\begin{document}

\section{Introduction}
The production of heavy quarkonium at hadron colliders provides a
useful laboratory for testing the high-energy limit of quantum
chromodynamics (QCD) as well as the interplay of perturbative and
nonperturbative phenomena in QCD. This high-energy processes are
dominated by the multi-Regge final states, when the contribution of
partonic subprocesses involving t-channel parton exchanges to the
production cross section
 can become dominant. Then the transverse momenta of the incoming partons and their off-shell properties can
 no longer be neglected, and we deal with "Reggeized" $t$-channel partons.

  We use gauge-invariant approach to introduce off-shell properties of the incoming partons to the matrix element of the
  hard subprocess, which is based on the gauge-invariant effective action for
  QCD in the Regge limit~\cite{Lipatov95, LipatovVyazovsky}. This effective field theory incorporates
   besides ordinary gluons and quarks, also fields of the Reggeized gluons and quarks.
    Corresponding Feynman rules where obtained in Refs.~\cite{LipatovVyazovsky,NLOBFKL,FeynRules},
    and they can be used to obtain amplitudes with off-shell incoming
    partons.

  \section{Heavy quarkonium production in the LO PRA.}
  In the present note, we study the process of heavy quarkonium production, exploiting the leading order (LO) of parton
  Reggeization approach (PRA) and non-relativistic QCD (NRQCD) factorization hypothesis \cite{NRQCD, Maltoni}. Working at the LO
   in  $\alpha_s$ and the relative velocity of heavy quarks $v$, we consider the following partonic subprocesses,
   which describe heavy quarkonium production at high energy:
\begin{eqnarray}
R(q_1) + R(q_2) &\to& {\cal H}
[{^3P}_J^{(1)},{^3S}_1^{(8)},{^1S}_0^{(8)},{^3P}_J^{(8)}](p),
\label{eq:RRtoH}\\
 R(q_1) + R(q_2) &\to& {\cal H} [{^3S}_1^{(1)}](p) + g(p'),
\label{eq:RRtoHG}
\end{eqnarray}
where $R$ is the Reggeized gluon and $g$ is the Yang-Mills gluon,
respectively, with four-momenta indicated in parentheses, ${\cal
H}[n]$ is the physical charmonium state, produced through the
intermediate Fock state of the $q\bar{q}$ pair
$[{}^{2S+1}L_J^{(1,8)}]$ with the spin $S$, total angular momentum
$J$, orbital angular momentum $L$ and in the color-singlet $^{(1)}$
(CS) or in the color-octet $^{(8)}$ (CO) quantum states. The squared
amplitudes $\overline{|{\cal A}(R + R \to q\bar
q[^{2S+1}L_J^{(1,8)}])|^2}$ of the relevant subprocesses in the LO
PRA have been obtained in the paper~\cite{KSVbottom}, and they are
shown to be coinciding with our earlier results \cite{KSVcharm} in
the $k_T$-factorization approach \cite{GLR}.

In the high-energy factorization, the hadronic cross section is
presented as convolution of off-shell partonic cross-section and
unintegrated parton distribution functions (uPDF). The best
agreement with experiment for the charmonium and bottomonium
production,
 as well as for other processes at low $p_T$ ~\cite{DY}, was obtained with
 the use of the KMR~\cite{KMR} procedure, which prescribes how to obtain the uPDFs from the conventional integrated ones.
 For the more details of the formalism, computational formulas and the fitting
procedure for the CO NMEs, see the Ref.~\cite{SNScharm} and
references therein.

\begin{figure}
\begin{tabular}{ccc}
\includegraphics[width=0.31\textwidth]{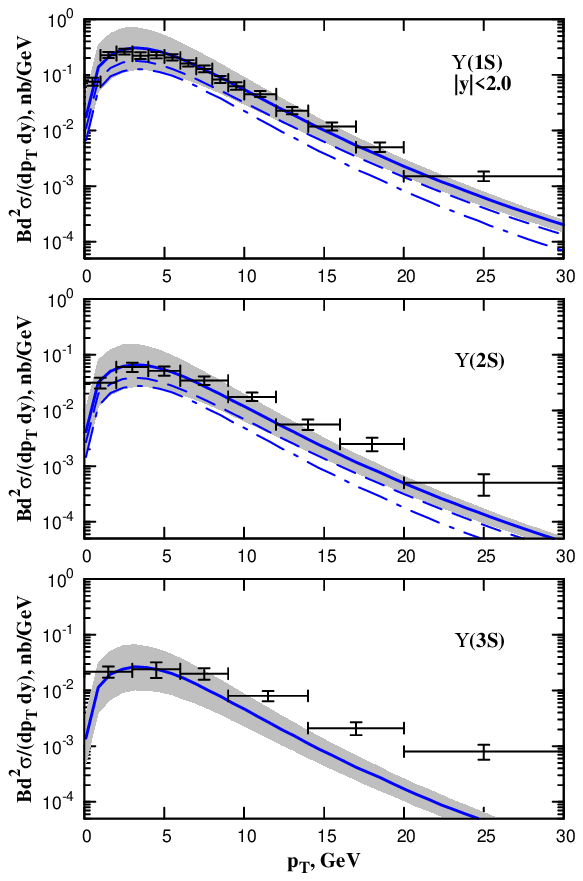} & \includegraphics[width=0.31\textwidth]{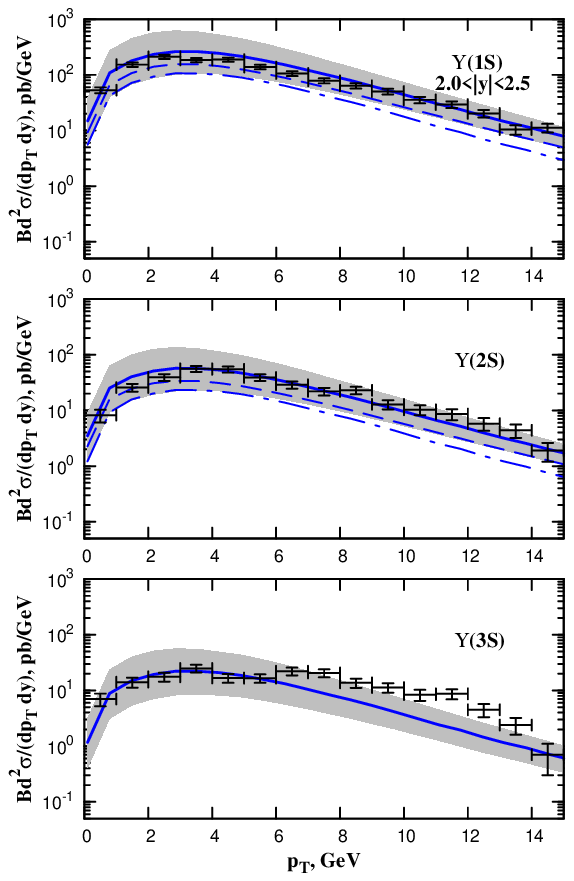}
& \includegraphics[width=0.31\textwidth]{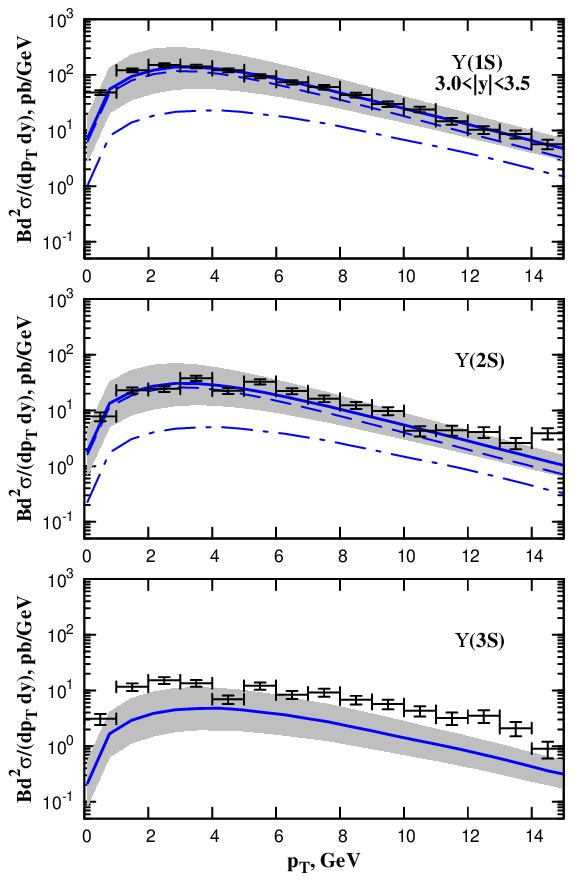}\\
(a) & (b) & (c)
\end{tabular}
\caption{Comparison of the LO PRA CS model predictions for the
inclusive $\Upsilon(nS)$ production with the CMS~\cite{CMSUpsi} (a),
 and LHCb~\cite{LHCbUpsi} (b, c) data. Dashed lines -- direct production, dash-dotted lines -- feddown decays contribution,
  solid lines -- their sum. The branching-fractions $B(\Upsilon(nS)\to\mu^+\mu^-)$ are included.}
\label{fig1}
\end{figure}
 \section{Numerical results.}
  Our latest numerical results for the prompt charmonium production in the Regge limit of QCD are described in the Ref.~\cite{SNScharm}.
   We have shown that  it is possible to describe the LHC experimental data on the prompt charmonium production
   at the $\sqrt{S}=7$ TeV in a wide kinematical range ($2<p_T<20$ GeV and $|y|<3.5$) with a good accuracy, using
    the CO NMEs extracted from  Tevatron data at the $\sqrt{S}=1.8$ TeV and $1.96$ TeV. The fitted CO NMEs are also shown to be compatible
     with NLO collinear parton model  results of Ref.~\cite{KniehlPSI}.

  Here we discuss our results for the inclusive $\Upsilon(nS)$ transverse-momentum spectra in the LO PRA and CS approximation of the NRQCD.
   As in the Ref.~\cite{SNScharm}, here we used the KMR prescription with the MRST-2006 \cite{MRST2006}
  set of the PDFs as the collinear input. The values of the CS NMEs were used the same as in the
  Ref.~\cite{KSVbottom},
  taking into account corrected values of branching fractions for the radiative decays of the $\Upsilon(nS)$ and $\chi_{bJ}(nP)$ states, accordingly
   the latest version of Particle Data Group \cite{PDG}.
  Due to the lack of space, we do not present in this note our new predictions for the old Tevatron
   data which show the qualitative features same with the features indicated below for the LHC data.

  In the Figure \ref{fig1}, the comparison of the LO PRA prediction on the $\Upsilon(nS)$ production through the CS mechanism with
  LHC experimental data for inclusive $\Upsilon(nS)$ production in the $pp$-collisions at the $\sqrt{S}=7$ TeV,
  measured by the CMS~\cite{CMSUpsi} and LHCb~\cite{LHCbUpsi} collaborations, is presented. The factorization and renormalization
   scales in this calculation are chosen to be $\mu_R=\mu_F=\xi M_T$ where $M_T=\sqrt{M^2+p_T^2}$ is the heavy quarkonium transverse mass, and $\xi$ is varied in the interval $1/2<\xi<2$ to estimate
   the scale-choosing uncertainty of our calculation, and the result of this variation is indicated on the plots by the gray band.

  Figure \ref{fig1} shows that the CS contributions are not sufficient to describe the data for the production of the
  $\Upsilon(2S)$ and $\Upsilon(3S)$ states, whereas the data on the $\Upsilon(1S)$ production in the region of $p_T<15$ GeV are
   described quite well. Moreover, description of $\Upsilon(1S)$ production in the rapidity interval $3<|y|<3.5$ becomes better,
    because of the decrease of the fraction of feeddown-decays contribution with rapidity. These features strongly indicate the
    necessity to include the CO intermediate state contributions to the calculations. As in the case of the charmonium
     production \cite{KSVcharm, SNScharm, KniehlPSI}, the leading contribution here will be from the CO $^3S_1^{(8)}$ intermediate
      state, which $p_T$-spectrum has the much lower slope, than the slope of the CS contribution.
      So, as a result of the fit, this contribution should affect the inclusive $\Upsilon(1S)$ $p_T$-spectrum only
      for the $p_T>15$ GeV, while it's contributions to the production of $\Upsilon(2S)$, $\Upsilon(3S)$ and $P$-wave states will be greater.
      The fit of the CO NMEs for bottomonium production will be the subject of our future study.

\end{document}